\documentclass[12pt]{article}
\usepackage{natnib}
\usepackage{times}
\usepackage{geometry}
\usepackage{graphicx}
\usepackage{apjfonts}
\usepackage[utf8]{inputenc}
\usepackage{enumitem,amssymb}
\usepackage{ragged2e}
\usepackage[colorlinks = true,linkcolor=blue,urlcolor=blue,citecolor=blue,anchorcolor=blue]{hyperref}
\newlist{thematic}{itemize}{8}
\setlist[thematic]{label=$\square$}
\usepackage{pifont}
\newcommand{\cmark}{\ding{51}}%
\newcommand{\done}{\rlap{$\square$}{\raisebox{2pt}{\large\hspace{1pt}\cmark}}%
\hspace{-2.5pt}}

\usepackage[superscript,biblabel]{cite}
\newcommand{\apj} {ApJ}

\newcommand{\aap} {A\&A}
\newcommand{\mnras} {MNRAS}
\newcommand{\apjl} {ApJ Letters}
\newcommand{\aapr} {A\&AR}
\newcommand{\aj} {AJ}
\newcommand{\apjs} {ApJS}
\newcommand{\pasp} {PASP}
\begin{document}
\raggedright
\huge
Astro2020 Science White Paper \linebreak

Science at the edges:\ internal kinematics of globular clusters' external fields \linebreak
\normalsize

\noindent \textbf{Thematic Areas:} \hspace*{60pt} $\square$ Planetary
Systems \hspace*{10pt} $\square$ Star and Planet
Formation \hspace*{20pt}\linebreak
$\square$ Formation and Evolution of Compact Objects \hspace*{31pt}
$\square$ Cosmology and Fundamental Physics \linebreak
$\square$ Stars and Stellar Evolution \hspace*{1pt} \done Resolved
Stellar Populations and their Environments \hspace*{40pt} \linebreak
$\square$ Galaxy Evolution \hspace*{45pt} $\square$ Multi-Messenger
Astronomy and Astrophysics \hspace*{65pt} \linebreak

\textbf{Principal Author:}
\linebreak
Name:	Andrea Bellini
 \linebreak						
Institution:  STScI
 \linebreak
Email: \href{mailto:bellini@stsci.edu}{bellini@stsci.edu}
 \linebreak
Phone: +1 (410) 338-4431

\medskip
\medskip
\begin{flushleft}
\hskip -2.0mm \begin{tabular}{ll}
\multicolumn{1}{l}{\textbf{Co-authors:} (names and institutions)}\\
Mattia Libralato, STScI             &  Sangeeta Malhotra, NASA/Goddard        \\
Jay Anderson, STScI                 &  Peter Melchior, Princeton              \\
David Bennett, NASA/Goddard         &  Ed Nelan, STScI                        \\
Annalisa Calamida, STScI            &  Jason Rhodes, JPL                      \\
Stefano Casertano, STScI            &  Robyn E. Sanderson, UPenn              \\
S. Michael Fall, STScI                 &  Michael Shao, JPL                      \\
B. Scott Gaudi, Ohio State University  &  S. Tony Sohn, STScI                    \\
Raja Guhathakurta, UCSC             &  Enrico Vesperini, Univesity of Indiana \\
Shirley Ho, Berkeley                &  Roeland P. van der Marel, STScI, JHU   \\
Jessica R. Lu, Berkeley             &\\
\end{tabular}
\end{flushleft}

\medskip

\medskip
~\\
\textbf{Abstract:} 

The outer regions of globular clusters can enable us to answer many
fundamental questions concerning issues ranging from the formation and
evolution of clusters and their multiple stellar populations to the
study of stars near and beyond the hydrogen-burning limit and to the
dynamics of the Milky Way. The outskirts of globular clusters are
still uncharted territories observationally. A very efficient way to
explore them is through high-precision proper motions of low-mass
stars over a large field of view. The Wide Field InfraRed Survey
Telescope (WFIRST) combines all these characteristics in a single
telescope, making it the best observational tool to uncover the wealth
of information contained in the clusters' outermost regions.

\pagebreak

\subsection*{Exploring uncharted territories}

\vskip -1mm
~~~Over the last two decades, our simple concept of globular clusters
(GCs) as ``spherical, kinematically-isotropic, non-rotating systems
composed of a single stellar population'' has radically changed, thanks to
exquisite spectroscopic, photometric and astrometric studies (the last two mainly
made possible by the \textit{Hubble Space Telescope},
\textit{HST}). The vast majority of these studies have been focusing
on the clusters' cores and surrounding central regions. As a
consequence, the outskirts of GCs and out to beyond their nominal tidal
radius formally remain ``uncharted territories'' from the observational
point of view.

\begin{figure}
    \centering
    \includegraphics[width=\textwidth]{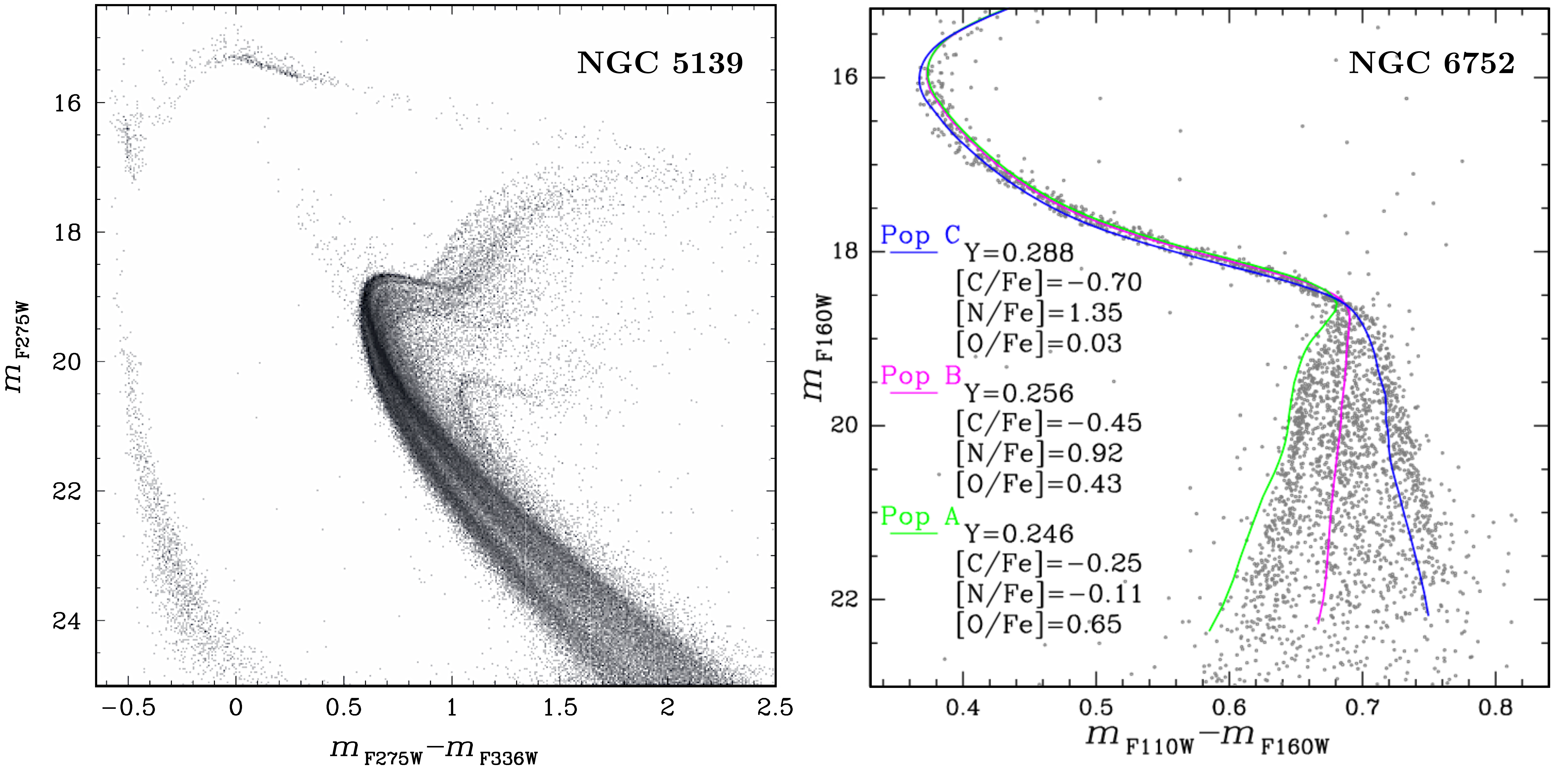}
\vskip -3mm
    \caption{\small \textit{(Left:)} UV CMD of NGC~5139 ($\omega$~Cen), from
      \citet{2017ApJ...842....6B}. \textit{Right:} IR CMD of NGC~6752,
      from \citet{2019MNRAS.484.4046M}. In both cases, multiple
      sequences are clearly visible on the CMDs. The UV filters F275W
      and F336W trace key molecular bands (in particular OH and NH)
      that help to differentiate between first- and second-generation stars. The IR color
      F110W$-$F160W ($J-H$) is particularly sensible to oxygen
      variations below the main-sequence knee.\looseness=-4}
    \label{f1}
\vskip -4mm
\end{figure}

\medskip
\textbf{Multiple stellar populations.~} Essentially all GCs studied
with high precision (e.g., \citealt[][and references
  therein]{2015AJ....149...91P}) show clear evidence of multiple
stellar populations (mPOPs). Stars of different populations form
distinct sequences on a color-magnitude diagram (CMD, e.g.,
\citealt{2017ApJ...844..164B}, see Fig.~\ref{f1}). These sequences can
also be traced to different abundances of light elements (such as Na,
O, Al, Mg observed spectroscopically; see e.g.,
\citealt{2012A&ARv..20...50G} and references therein), helium (which
is hard to observe spectroscopically, \citealt{2011A&A...531A..35P,
  2011ApJ...728..155D, 2018MNRAS.481.5098M}), and iron (e.g.,
\citealt{2009A&A...505.1099M}). First stellar generations have a
composition typical of the proto-galactic interstellar matter out of
which they formed. Subsequent generations  have increasingly
higher helium, N and Na, and are depleted in C and O. mPOPs are a
ubiquitous feature of GCs, but no two clusters are alike
(\citealt{2015MNRAS.454.4197R}).\looseness=-2

\medskip
~~~While photometry and spectroscopy have begun to shed some light on the
mPOP phenomenon in terms of chemical properties, multiplicity of
distinct sequences and ranges of ages, many questions remain still
unanswered, e.g.:\ What sequence of events led to the formation of
mPOPs? How did formation processes and dynamical evolution shape
today's GCs?  The structural and kinematic properties of GCs are two
additional key pieces of information to help us build a complete
picture of the formation and evolution of GCs and their mPOPs. In this
respect, the measurement of stellar proper motions (PMs) represents
\textit{a very effective way} to constrain the structure, formation
and dynamical evolution of these ancient stellar systems and, in turn,
that of the Milky Way itself.\looseness=-2

\medskip
~~~Some mPOP formation scenarios propose second-generation stars to
form more centrally concentrated than first-generation stars (e.g.,
\citealt{2008MNRAS.391..825D}). As GCs evolve over time, two-body
relaxation processes tend to progressively erase this initial spatial
segregation, starting from the clusters' cores and moving towards the
outskirts. The GCs' core relaxation time is short---a few $10^7-10^8$
yr---compared to the age of GCs of $12-13\times 10^9$ yr, so first- and
second-generation stars are now fully mixed and isotropic in the
central regions.  Numerical simulations predict that the outer regions
of massive GCs, characterized by a longer two-body-relaxation time,
should still retain fossil memory of the initial segregation of
second-generation stars (e.g., \citealt{2013MNRAS.429.1913V}). This is
indeed what has been observed in some GCs (e.g.,
\citealt{2009A&A...507.1393B, 2013ApJ...765...32B,
  2016MNRAS.463..449S}).

\medskip
~~~Kinematically, stars in GCs are expected to diffuse outwards
preferentially along radial orbits, resulting in an increased radial
anisotropy of their velocity-dispersion profile. Over many two-body
relaxation times stars tend to have a more isotropic velocity profile.
Second-generation stars were born at a later stage (and more centrally
segregated) than first-generation stars, so that we expect them to
show a higher degree of velocity anisotropy than first-generation
stars. The velocity-dispersion anisotropy can be directly measured via
PMs alone as the ratio between the tangential ($\sigma_{\rm tan}$) and
the radial ($\sigma_{\rm rad}$) components of the velocity dispersion,
and cannot be done by just using line-of-sight velocities, since two
directions of motion are needed. The outskirts of GCs represent the
best place to still find second-generation stars more radially
anisotropic than first-generation stars (e.g.,
\citealt{2013ApJ...771L..15R, 2015ApJ...810L..13B,
  2018ApJ...853...86B}).  Furthermore, pinpointing the radial distance
at which the isotropy-to-radial-anisotropy transition happens can help
us set tight constrains on the time evolution of mPOPs and the
dynamical history of GCs.

\medskip
\textbf{GCs at the edge (and beyond).~} While mPOPs are now the
hottest topic in GC studies, there are several other open key
questions regarding GCs as a whole that could potentially be
substantially clarified with a thorough analysis of the clusters'
outer regions.
\begin{itemize}
\item{\vskip -2mm \textbf{What is the role of angular momentum in
    cluster formation and evolution?} A large number of GCs exhibits a
  significant degree of systemic rotation (e.g.,
  \citealt{2012A&A...538A..18B, 2014ApJ...787L..26F,
    2018MNRAS.481.2125B, 2018ApJ...860...50F,
    2018MNRAS.473.5591K}). The overwhelming majority of these studies
  is based on line-of-sight (LOS) measurements. A combination of LOS
  measurements with PMs will allow us to derive the three-dimensional
  kinematic profile (see left panel in Fig.~\ref{f2}), which
  can be used to shade light on this important topic (e.g.,
  \citealt{2017ApJ...844..167B}).}
\item{\vskip -2mm \textbf{What is the current state of energy
    equipartition in clusters?}  For over forty years, GCs were
  expected to evolve towards a state of complete energy equipartition
  over many two-body relaxation times, so that the stellar velocity
  dispersion $\sigma_{\rm vel}$ scales with stellar mass $m$ as
  $\sigma_{\rm vel} \propto m^{-\eta}$, with $\eta = 0.5$
  (\citealt{1969ApJ...158L.139S, 1987degc.book.....S}). However,
  recent simulations (e.g., \citealt{2013MNRAS.435.3272T,
    2016MNRAS.458.3644B, 2017MNRAS.464.1977W}) show that GCs can only
  reach partial energy equipartition, and that the value of $\eta$
  depends on the distance from the cluster's center and on stellar
  mass. Measuring the detailed state of energy equipartition of a GC
  requires high-precision PM measurements for a large number of stars
  over a wide range of stellar masses (e.g.,
  \citealt{2018ApJ...853...86B}) and radial distances (e.g.,
  \citealt{2018ApJ...861...99L}:\ see the second panel of
  Fig.~\ref{f2}).\looseness=-2}
\item{\vskip -2mm \textbf{What can we learn from stars near and beyond
    the hydrogen-burning limit?}  There exists a minimum mass below
  which a star cannot ignite thermonuclear burning of hydrogen. This
  hydrogen-burning limit (HBL) separates main-sequence stars from
  \textbf{brown dwarfs}, which are characterized by a large difference
  in luminosity for a small variation in mass. The best place to
  observe stars near (and beyond) the HBL is in the outskirts of GCs,
  where crowding is low and low-mass stars are the majority. In
  addition, stars in GCs are all at the same distance and age from us,
  so that a difference in luminosity directly translates into a
  difference in mass. Because GCs are old, stars with masses below the
  HBL had time to fade away from those above the limit, creating a
  clear gap in the luminosity function (e.g., \citealt
  {2001ApJ...560L..75B, 2016ApJ...817...48D}, see the third panel of
  Fig.~\ref{f2}) that can be use, e.g., to provide an independent age
  estimate for the clusters. Proper motions are needed to identify
  bona-fide cluster members near the HBL from background unresolved
  galaxies.}
\item{\vskip -2mm \textbf{What is the role of the Galactic tidal
      field?} During their evolution, GCs tend to become radially
    anisotropic the farther from their cores. In the outermost cluster
    regions, where the interaction of cluster stars with the Galactic
    tidal field become significant, stars with more radial orbits are
    preferentially lost in the field. As a result, the velocity field
    of surviving stars is expected to become isotropic or even
    slightly tangentially anisotropic (see, e.g.,
    \citealt{2000MNRAS.316..671T, 2003MNRAS.340..227B,
      2014MNRAS.443L..79V, 2016MNRAS.455.3693T}, right panel of
    Fig.~\ref{f2}).  The interplay between the Galactic tidal field
    and the stellar orbits in a GC is still poorly constrained
    observationally, and it is not clear how deep into the cluster's
    potential well the influence of the tidal field can be.}
\begin{figure}
    \centering
    \includegraphics[width=\textwidth]{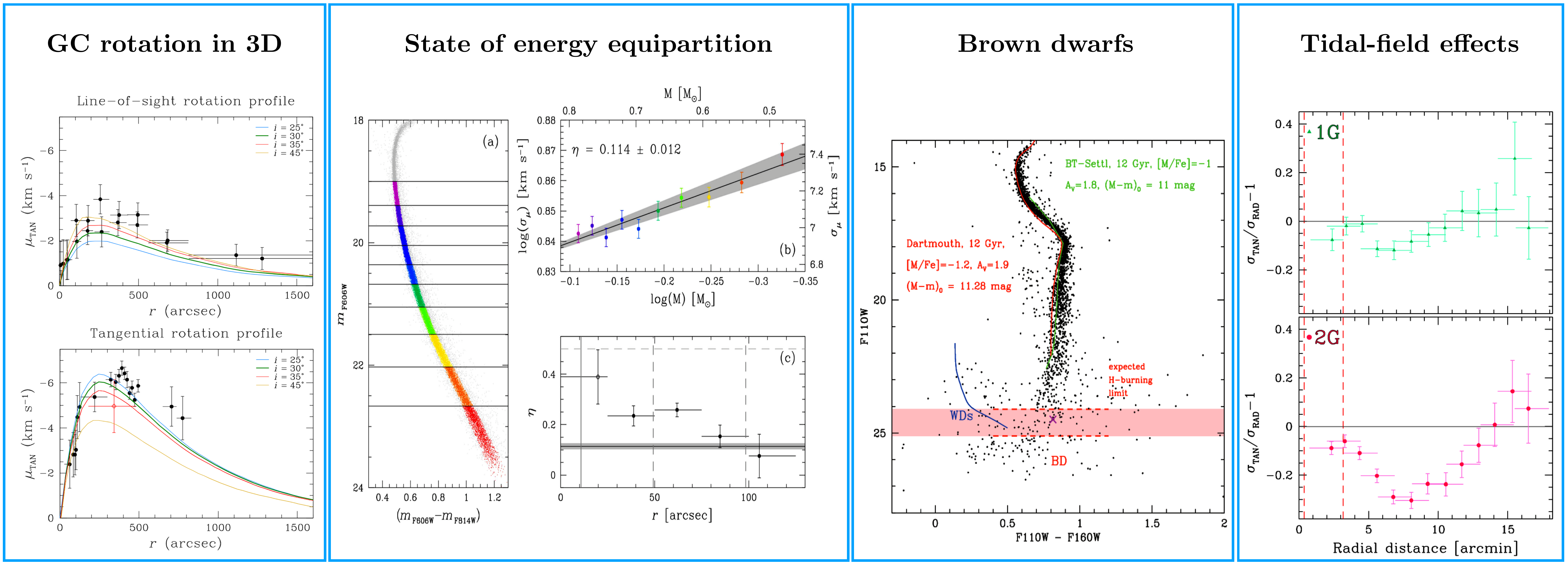}
\vskip -3mm
    \caption{\small \textit{(From left to right:)} (1) The systemic
      rotation of the GC 47~Tuc (from
      \citealt{2017ApJ...844..167B}). (2) Measure of the state of
      energy equipartition in the GC NGC~362 (from
      \citealt{2018ApJ...861...99L}). The faint limit of the CMD of
      the GC M~4 near the HBL (from
      \citealt{2016ApJ...817...48D}). (4) Anisotropy radial profiles
      of first- and second-generation stars in the GC 47~Tuc (from
      \citealt{2018MNRAS.479.5005M}).}
    \label{f2}
\vskip -4mm
\end{figure}
\item{\vskip -2mm \textbf{Tidal tails as a powerful dynamical tool.}
  The internal dynamical evolution of GCs, driven by two-body
  relaxation along with the effects of disk and bulge tidal shocks,
  causes cluster stars to escape, forming tidal tails. These tidal
  tails are present as departures in the surface density profiles at
  large radii, with a break from a King profile
  (\citealt{1966AJ.....71...64K}) at the tidal radius followed by a
  power-law-like decline which varies from cluster to cluster (but see
  also the role of potential escapers, e.g.,
  \citealt{2019arXiv190108072D}). Evidence of tidal tails has been
  found in several GCs (see, e.g., \citealt{1995AJ....109.2553G,
    2001ApJ...548L.165O, 2003AJ....126.2385O, 2006ApJ...639L..17G,
    2014A&A...572A..30K, 2019arXiv190209544I}), implying this might be
  a common feature. The identification and analysis of these tails,
  together with their shape, extension, orientation and stellar
  content, provides a wealth of information on the cluster dynamical
  evolution, on the evolution of its stellar-population content and
  mass function, on the cluster's orbit and on the Galactic potential
  (see, e.g., \citealt{2003gmbp.book.....H, 2008gady.book.....B}), and
  on the possible fingerprints of dark matter substructures
  (\citealt{2018arXiv181103631B}).  In addition, some mPOP formation
  models (e.g., \citealt{2008MNRAS.391..825D}) predict GCs to have
  been much more massive at birth, with most of their first-generation
  stars now lost into the field. Measuring the population ratio of
  extra-tidal stars would provide critical elements to prove or
  dismiss these theories.  Extra-tidal stars are rare, and the
  most-obvious places to look for them are the immediate surroundings
  of GCs, just outside their tidal radius along the direction of
  motion of the clusters.\looseness=-2}
\end{itemize}

\subsection*{\vskip -8mm The path forward}
\vskip -1mm ~~~\textbf{Observational requirements.} All the science
cases listed above share specific needs:
\begin{enumerate}
\item{\vskip -1mm \textbf{Wide field of view}.  The tidal radius of GCs
  ranges from several arcmin up to about 1 degree, and tidal tails
  obviously extend further out, so {\em the wider the field of view of
    a telescope, the more efficient and less time consuming
    observations will be.}}
\item{\vskip -3mm \textbf{Deep, near-IR.} Because of the effects of
  mass segregation, bright, massive stars are more centrally
  concentrated, while the cluster outskirts are dominated by faint,
  less massive stars. Because of stellar evolution, stars spend about
  99\% of their life time as (faint) MS stars, and the remaining 1\%
  as (bright) evolved stars. It follows that the vast majority of
  stars in the outskirts of GCs are faint, low-mass stars. Low-mass
  stars emit most of their light at redder wavelength than their
  brighter counterparts, especially at the faint end, so that {\em an
    efficient observational strategy would employ near-IR} rather than
  optical or UV filters.}
\item{\vskip -3mm \textbf{Cluster-field decontamination.} The stellar
  density in GCs quickly drops moving outwards from the core, and soon
  foreground and background sources become the dominant sources of
  contamination. With this respect, {\em PMs represent the best tool
    to clearly isolate cluster members from the field population.}}
\item{\vskip -3mm \textbf{Astrometric precision.} Typical velocity
  dispersions of stars in the outskirts of GCs are of the order of a
  just a few km\,s$^{-1}$. Proper-motion errors need to be subtracted
  in quadrature from the observed velocity dispersions to obtain the
  stellar intrinsic velocity dispersion, and typically astrometrists
  discard PM measurements if their error is larger than half that the
  intrinsic velocity dispersion (e.g., \citealt{2014ApJ...797..115B,
    2015ApJ...803...29W, 2018ApJ...861...99L}). At the typical GC
  distance of 10 kpc, a $\lesssim$1 km\,s$^{-1}$ PM relative precision
  translates into an angular precision of $\lesssim$20
  $\mu$as\,yr$^{-1}$.}
\end{enumerate}

\begin{figure}
    \centering
    \includegraphics[width=\textwidth]{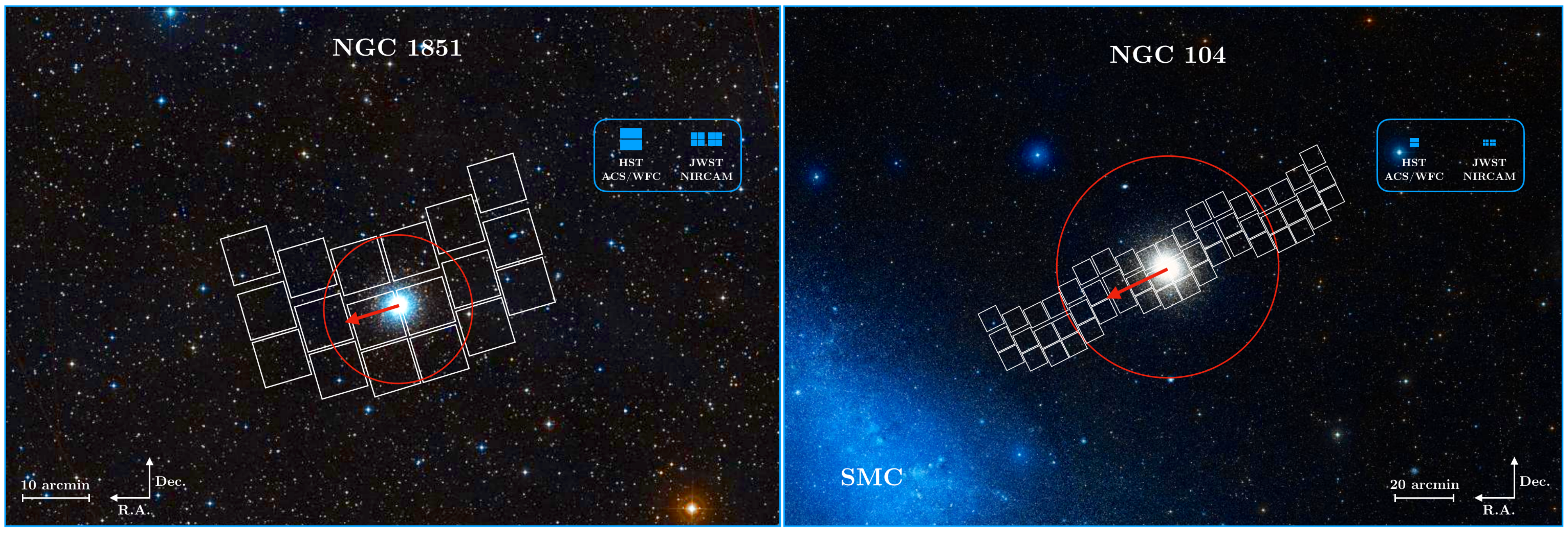}
\vskip -3mm
    \caption{\small A typical-size GC (\textit{left}) and one of the
      largest GCs (\textit{right}) can easily be observed with just
      one or a few WFIRST pointings. The \textit{HST} and
      \textit{JWST} field of views are also shown, for comparison. In
      both panels, the red circle shows the nominal tidal radius of
      the clusters. Observations outside the tidal radius along the
      direction of motion of the clusters (red arrow) allows the study
      of tidal tails and the effects of the tidal field.\looseness=-4}
    \label{f3}
\vskip -4mm
\end{figure}

\vskip -1mm ~~~\textbf{\textit{Gaia}'s end-of-mission PMs of around 20
  $\mu$as\,yr$^{-1}$ are expected to be limited in magnitude to stars
brighter than $G$$\sim$17} (\citealt{2017MNRAS.467..412P}). This
limit implies that only the few evolved stars present in the clusters'
outskirts would have PM measurements of the required precision, thus
severely limiting the available statistics and make most of the
proposed investigations impossible. Even for the few closest GCs for
which \textit{Gaia}'s PMs are of adequate precision down to $\sim$1 mag
below the turn off, the range in mass of these stars is too narrow to
derive meaningful values of $\eta$.

\medskip
~~~\textbf{\textit{HST} has a pencil-beam field-of-view.}  To map the
outer regions of typical GCs with the required precision would imply
the need for hundreds of \textit{HST} orbits in each of at least two
epochs, and surveying several GCs in this way would simply be
unfeasible (and unreasonable). \textit{JWST} will have a similar
astrometric performance and suffer from the same field-of-view
limitations as \textit{HST}.

\medskip
~~~\textbf{Ground-based telescopes are not precise enough}.
Atmospheric effects and telescope flexures make their point-spread
function and geometric distortion very unstable at the submas
level. As a result, ground-based PM measurements typically have
precisions of a few tenths of a mas\,yr$^{-1}$ at best (e.g.,
\citealt{2006A&A...454.1029A, 2009A&A...493..959B,
  2010A&A...513A..50B, 2014A&A...563A..80L,2015MNRAS.450.1664L}):\ at
least an order of magnitude too large of what is need by the proposed
studies. Current and near-future adaptive-optic-equipped (AO, e.g.,
Keck, E-ELT, US-ELTs) detectors can in principle have a better
astrometric performance than \textit{HST}, but their field-of-view
is/will be even smaller than that of \textit{HST}.

\medskip
~~~\textbf{WFIRST represents the perfect tools for the proposed
  investigations}: it combines the astrometric capabilities of
\textit{HST} with a single-shot spatial coverage typical of that of
wide-field ground-based telescopes (see \citealt{2017arXiv171205420T}
for a detailed description of the expected WFIRST astrometric
performance). Most GCs fit within one WFIRST pointing, and even the
largest GCs can be observed with just two or three pointings
(Fig.~\ref{f3}). Assuming a conservative 0.01 pixel position precision
per single WFIRST image (as it is the case for \textit{HST}, e.g.,
\citealt{2011PASP..123..622B}), a 20 $\mu$as\,yr$^{-1}$ precision can
be obtained in four years (the nominal mission lifetime) with about
200 exposures (or just 100 exposures if the mission is extended two
more years). If the single-exposure precision can be improved by a
factor of two (or more)---which is not unreasonable given the much
more stable thermal environment of WFIRST's orbit compared to that of
\textit{HST}'s low-Earth orbit---than these figures also scale down by
the same factor, making what would otherwise be a hard or even
impossible observing strategy relatively easy to achieve.

\end{document}